\documentstyle[12pt]{article}

\title{\textbf{Non Gaussianity of General Multiple-Field Inflationary Models}}
\author{\textbf{Cheng-Yi Sun\footnote{cysun@mails.gucas.ac.cn}
\ and \ De-Hai Zhang\footnote{dhzhang@gucas.ac.cn}}\\
Department of Physics,\\
The Graduate School of The Chinese Academy of Sciences,\\
Beijing 10049, P.R.China.}

\begin{document}
\maketitle
\begin{abstract}
Using the ``$\delta N$-formalism", We obtain the expression of the
non-Gaussianity of multiple-field inflationary models with the
nontrivial field-space metric. Further, we rewritten the result by
using the slow-rolling approximation.
\end{abstract}



\section{Introduction}
In modern cosmology, the inflation paradigm plays an important role.
The simplest classes of inflation models predict
Gaussian-distributed perturbations and a nearly scale-invariant
spectrum of the primordial density perturbations \cite{CISL}. This
is in good agreement with cosmological observations \cite{WMAP}.
Despite the appealing simplicity behind the central idea of
inflation, it has proved difficult to discriminate between the large
number of different models that have been developed to date
\cite{p9807278}. However, it is believed that the deviation away
from the Gaussian statistics represents a potential powerful
discriminant between the competing inflationary models. At the same
time, it is necessary to extend the theoretical framework beyond the
leading-order effects of scale-invariant, Gaussian fluctuations, in
order to understand the early universe further. So, recently, the
non-Gaussianity has attracted considerable interest. (See
\cite{a0406398} for a review.) In \cite{a0210603}, Maldacena gave a
general analysis of non-Gaussian perturbations in single field
inflationary models. His result is that the bispectrum of the
curvature perturbation for squeezed triangles ($k_1\ll k_2,k_3$) is
proportional to the tilt of the primordial power spectrum, and hence
is small. Then it seems that only multiple-field inflationary models
is like to generate significant non-Gaussian perturbations
\cite{a0504508,a0506056,a0506704,a0509719,a0504045,a0504046}. In
\cite{a0504045}, Lyth and Rodriguez have shown that the
non-Gaussianity of the curvature perturbation in multiple field
models can be simply expressed in the so-called ``$\delta
N$-formalism" \cite{a0411220}. There the ``separate universe"
approach \cite{a0306620} has been used to define the curvature
perturbation. It should be noted that this approach is valid only
for perturbations on super-Hubble scales. But Lyth and Rodriguez
only express $f_{NL}$ on a field space with the trivial metric,
where $f_{NL}$ is the non-linear parameter. In \cite{a0506056}, the
authors have given the expression of $f_{NL}$ involving the metric
of the field space, $G_{IJ}$, explicitly. But they restricted their
attentions only on the metric that \emph{can be brought to the
field-independent form $G_{IJ}=\delta_{IJ}$ by an appropriate choice
of parametrization.}

In this paper, first, the result in \cite{a0504045} is generalized
to the case with a generic field-space metric. It is found that
the generalized expression is similar to the expression obtained
in \cite{a0506056} for a trivial field-space metric. Then this
expression is rewritten in terms of slow-rolling parameters.

\section{The background and the curvature perturbation}
The starting point is the effective action of the simple coupling
system of Einstein gravity and scalar fields with an arbitrary
inflation potential $V(\varphi^I)$
\begin{equation}
\label{action}S=\int{\sqrt{-g}d^4x[\frac{M_p^2}{2}R-\frac{1}{2}
G_{IJ}\partial_\mu\varphi^I\partial^\mu\varphi^J-V(\varphi^I)]},
\end{equation}
where $G_{IJ}\equiv G_{IJ}(\varphi^K)$ represents the metric on the
manifold parameterized by the scalar field values, the 'target
space' metric, and $8\pi G=M_p^{-2}$ represents the reduced Planck
mass. Units are chosen such that $c=\hbar=1$. For the background
model, the Friedmann-Robertson-Walker metric is used,
\begin{equation}
\label{FRWmetric}ds^2=-dt^2+a^2(t)\delta_{ij}dx^idx^j.
\end{equation}
Take the background scalar fields as $\varphi^I(t)$. Then
background equations of the scalar fields are
\begin{equation}
\label{BgEqOfScalrF}\ddot{\varphi}^I+3H\dot{\varphi}^I+
\Gamma^I_{JK}\dot{\varphi}^J\dot{\varphi}^K+G^{IJ}V_{,I}=0,
\end{equation}
where
$\Gamma^I_{JK}=\frac{1}{2}G^{IL}(G_{JL,K}+G_{LK,J}-G_{JK,L})$, are
the target space Christoffel symbols. $H=\dot{a}/a$ is the Hubble
parameter, $\dot{\varphi}^I=d\varphi^I/dt$,
$\ddot{\varphi}^I=\frac{d^2\varphi^I}{dt^2}$ and
$V_{,I}=\frac{\partial V}{\partial\varphi^I}$,
$G_{IJ,K}=\frac{\partial G_{IJ}}{\partial\varphi^K}$. Basing on
the Einstein equation, we get
\begin{equation}
\label{EinsteinEq}\frac{\ddot{a}}{a}=-\frac{1}{3M_p^2}
(G_{IJ}\dot{\varphi}^I\dot{\varphi}^J-V).
\end{equation}
Together with the Friedmann equation
\begin{equation}
\label{FriedmannEq}H^2=\frac{1}{3M_p^2}
(\frac{1}{2}G_{IJ}\dot{\varphi}^I\dot{\varphi}^J+V),
\end{equation}
we get
\begin{equation}
\label{dHubbleDT}\dot{H}\equiv\frac{dH}{dt}=-\frac{1}{2M_p^2}
G_{IJ}\dot{\varphi}^I\dot{\varphi}^J.
\end{equation}

Now, let's consider the perturbed scalar fields as
$\varphi^I(t)+\delta\varphi^I(t,\textbf{x})$, and define the
curvature perturbation. Here the curvature perturbation refers to
the uniform density curvature perturbation, $\zeta$, which is still
equivalent to the comoving curvature perturbation on super horizon
scales in multiple field inflationary models. The curvature
perturbation is defined as the difference between an initial
space-flat fixed-$t$ slice and a final uniform energy density
fixed-$t$ slice (see \cite{Starobinsky,a0504045,a9507001} for
details),
\begin{equation}
\label{curvaturePer}\zeta(t,\textbf{x})=\delta N=H\delta t,
\end{equation}
where $N=\int{Hdt}$ is the integrated number of e-folds. Following
the argument in \cite{a0504045}, we expand the curvature
perturbation to the second order,
\begin{equation}
\label{deltaN}\zeta\simeq N_{,I}(t)\delta\varphi^I(\textbf{x})+
\frac{1}{2}N_{,IJ}(t)\delta\varphi^I(\textbf{x})\delta\varphi^J(\textbf{x}),
\end{equation}
where $N_{,I}=\frac{\partial N}{\partial\varphi^I}$,
$N_{,IJ}=\frac{\partial^2 N}{\partial\varphi^I\partial\varphi^J}$.
In this equation, it is the partial differentiation, not the
covariant differentiation that is used, which is the same as in
\cite{a0506056}. This is due to the definition of the curvature
perturbation.

\section{the non-linear parameter, $f_{NL}$}
The non Gaussianity of the curvature is expressed in the form
\begin{equation}
\label{defNonLinear}\zeta=\zeta_g-\frac{3}{5}f_{NL}(\zeta_g^2-\langle\zeta_g^2\rangle),
\end{equation}
where $\zeta_g$ is Gaussian, with $\langle\zeta_g\rangle=0$, and
$f_{NL} is the non-linear parameter$. For a generic cosmological
perturbation, $\psi(t,\textbf{x})$, we define its Fourier components
as
$\psi(\textbf{k})=\int{d^3x\psi(t,\textbf{x})e^{i\textbf{k}\cdot\textbf{x}}}$.
Then, using Eq.(\ref{defNonLinear}), we get
\begin{equation}
\label{FourierZetaFNL}\zeta(\textbf{k})=\zeta_g(\textbf{k})
-\frac{3}{5}f_{NL}\{\int{\frac{d^3k_1}{(2\pi)^3}[\zeta_g(\textbf{k}_1)
\zeta_g(\textbf{k}-\textbf{k}_1)]}-(2\pi)^3\delta^3(\textbf{k})\langle\zeta_g^2\rangle\}
\end{equation}
On the other hand, using Eq.(\ref{deltaN}), we get
\begin{equation}
\label{FourierZetaDeltaN}\zeta(\textbf{k})=N_{,I}\delta\varphi^I(\textbf{k})+
\frac{1}{2}N_{,IJ}\{\int{\frac{d^3k_1}{(2\pi)^3}
[\delta\varphi^I(\textbf{k}_1)\delta\varphi^J(\textbf{k}-\textbf{k}_1)]}-
(2\pi)^3\delta^3(\textbf{k})\langle\delta\varphi^I\delta\varphi^J\rangle\}.
\end{equation}
Here, in order to keep that $\langle\zeta\rangle=0$, we have added
the term, $-N_{IJ}\delta\varphi^I\delta\varphi^J$, to the
left-hand side of Eq.(\ref{deltaN}). And we have supposed that
$\langle\zeta_g^2\rangle$ and
$\langle\delta\varphi^I\delta\varphi^J\rangle$ are independent of
the spatial coordinates.

The spectrum of $\zeta_g$, $P_{\zeta}(k)$, is defined in the
standard way by
\begin{equation}
\label{spectrumOfZetaG}\langle\zeta_g(\textbf{k}_1)\zeta_g(\textbf{k}_2)\rangle
=(2\pi)^3\delta^3(\textbf{k}_1+\textbf{k}_2)P_{\zeta}(k_1),
\end{equation}
with $k\equiv|\textbf{k}|$. Together with
Eq.(\ref{FourierZetaFNL}), to first order of $f_{NL}$, the
spectrum of $\zeta$ is
\begin{equation}
\label{spectrumOfZetaFNL}\langle\zeta(\textbf{k}_1)\zeta(\textbf{k}_2)\rangle\simeq
\langle\zeta_g(\textbf{k}_1)\zeta_g(\textbf{k}_2)\rangle
=(2\pi)^3\delta^3(\textbf{k}_1+\textbf{k}_2)P_{\zeta}(k_1).
\end{equation}

For this multiple field model, by assuming the quasi exponential
inflation, basing on the results in \cite{a9507001,g9502002}, we
may define the spectrum of the scalar fields as
\begin{equation}
\label{spectrumOfField}\langle\delta\varphi^I(\textbf{k}_1)
\delta\varphi^J(\textbf{k}_2)\rangle
=(2\pi)^3\delta^3(\textbf{k}_1+\textbf{k}_2)P_{\delta\varphi}(k_1)G^{IJ}(\varphi_\ast),
\end{equation}
with
$\frac{k^3}{2\pi^2}P_{\delta\varphi}(k)=(\frac{H_\ast}{2\pi})^2$.
The subscript, $\ast$, means the value calculated at the moment
that the corresponding scale crosses out the Hubble horizon,
$k=aH$. Using the equations (\ref{FourierZetaDeltaN}) and
(\ref{spectrumOfField}), to leading order, we get the other
expression of the spectrum of $\zeta$,
\begin{equation}
\label{spectrumOfZetaDeN}\langle\zeta(\textbf{k}_1)\zeta(\textbf{k}_2)\rangle\simeq
(2\pi)^3\delta^3(\textbf{k}_1+\textbf{k}_2)P_{\delta\varphi}(k_1)N_{,I}N_{,J}G^{IJ}.
\end{equation}

Comparing Eq.(\ref{spectrumOfZetaFNL}) with
Eq.(\ref{spectrumOfZetaDeN}), we obtain the relation between
$P_{\zeta}(k)$ and $P_{\delta\varphi}(k)$,
\begin{equation}
\label{specZetaAndVarphi}P_{\zeta}(k)=P_{\delta\varphi}(k)N_{,I}N_{,J}G^{IJ}.
\end{equation}

Now let's calculate the bispectrum of $\zeta$. Using
Eq.(\ref{FourierZetaFNL}), to the first order of $f_{NL}$, we get
\begin{equation}
\label{bispectrumFNL}\langle\zeta(\textbf{k}_1)\zeta(\textbf{k}_2)\zeta(\textbf{k}_3)\rangle
=(2\pi)^3\delta^3(\textbf{k}_1+\textbf{k}_2+\textbf{k}_3)
\times\{-\frac{6}{5}[P_{\zeta}(k_1)P_{\zeta}(k_2)+cyclic]\},
\end{equation}
where $cyclic$ refers to the term,
$P_{\zeta}(k_2)P_{\zeta}(k_3)+P_{\zeta}(k_3)P_{\zeta}(k_1)$.
Above, we have used Eq.(\ref{spectrumOfZetaG}). On the other side,
using Eq.(\ref{FourierZetaDeltaN}), we get
\begin{equation}
\label{bispectrumDeN}\langle\zeta(\textbf{k}_1)\zeta(\textbf{k}_2)\zeta(\textbf{k}_3)\rangle
\simeq(2\pi)^3\delta^3(\textbf{k}_1+\textbf{k}_2+\textbf{k}_3)B(k_1,k_2,k_3),
\end{equation}
with
\begin{equation}
\label{B}B(k_1,k_2,k_3)\equiv
N_{,I}N_{,J}N_{,KL}G^{IK}G^{JL}[P_{\delta\varphi}(k_1)P_{\delta\varphi}(k_2)+cyclic],
\end{equation}
where $cyclic$ refers to the term,
$P_{\delta\varphi}(k_2)P_{\delta\varphi}(k_3)+P_{\delta\varphi}(k_3)P_{\delta\varphi}(k_1)$.
Here, we note that, in Eq.(\ref{bispectrumDeN}), we have ignored
the contribution of the term,
\begin{equation}
\label{intrinsicNG}N_{,I}N_{,J}N_{,K}\langle\delta\varphi^I(\textbf{k}_1)
\delta\varphi^J(\textbf{k}_2)\delta\varphi^K(\textbf{k}_3)\rangle,
\end{equation}
which comes from the intrinsic non Gaussianity of
$\delta\varphi^I(\textbf{k})$. We know, that for the case with the
trivial target space metric, $G_{IJ}=\delta_{IJ}$, in
\cite{a0507608}, it has been proved that the contribution of the
intrinsic non Gaussianity is small enough to be neglected. In this
paper, we suppose that, for nearly Gaussian perturbations,
$\delta\varphi^I$, the intrinsic non Gaussianity
(\ref{intrinsicNG}) is still small enough to be neglected in the
context of slow-roll inflation, and the bispectrum of $\zeta$ can
be obtained from Eq.(\ref{bispectrumDeN}).

Comparing Eq.(\ref{bispectrumFNL}) and Eq.(\ref{bispectrumDeN}),
we get the non-linear parameter as
\begin{equation}
\label{FNL}f_{NL}=-\frac{5}{6}\times\frac{G^{IM}G^{KN}N_{,I}N_{,K}N_{,MN}}
{(N_{,I}N_{,J}G^{IJ})^2}.
\end{equation}
This is an important result of this paper. Although this
expression is the same as the second term on the right-hand side
of Eq.(38) in Ref.\cite{a0506056}, here we obtain it for general
multi-field inflationary models.

\section{$f_{NL}$ and slow-rolling parameters}
In this section, with the slow-rolling condition, we try to express
$f_{NL}$ in term of the slow-rolling parameters. So we define some
parameters. The first is $\varepsilon$ defined as
\begin{equation}
\label{epsilonDef}\varepsilon=-\frac{\dot{H}}{H^2}.
\end{equation}
Using Eq.(\ref{dHubbleDT}), we get
\begin{equation}
\label{epsilonPhi}\varepsilon=\frac{G_{IJ}\dot{\varphi}^I\dot{\varphi}^J}{2M_p^2H^2}.
\end{equation}
Now let's use the slow-rolling approximation. Then
Eq.(\ref{BgEqOfScalrF}) becomes
\begin{equation}
\label{slowRoBgEqOfPhi}3H\dot{\varphi}^I+G^{IJ}V_{,J}\simeq0\Rightarrow
\dot{\varphi}^I\simeq-\frac{G^{IJ}V_{,J}}{3H}
\end{equation}
And Eq.(\ref{FriedmannEq}) becomes
\begin{equation}
\label{HubbleAndPote}H^2\simeq\frac{1}{3M_p^2}V.
\end{equation}
So $\varepsilon$
can rewritten approximately
as
\begin{equation}
\varepsilon
\simeq\frac{G^{IJ}V_{,I}V_{,J}M_p^2}{2V^2}.\label{epsilonV}
\end{equation}
Then we define another parameter, $\varepsilon_I$, as
\begin{equation}
\label{epsilonLIV}\varepsilon_I\equiv-\frac{V_{,I}M_p}{\sqrt{2}V}.
\end{equation}
This implies an relation,
$\varepsilon=G_{IJ}\varepsilon^I\varepsilon^J$.

In order to express $f_{NL}$ by the slow-rolling parameters, we
should firstly get the expression of $N_{,I}$. From
Eq.(\ref{curvaturePer}), we get
\begin{equation}
\label{deltaNHubble}\delta N=H\delta t=-\frac{1}{\varepsilon}d\ln
H\simeq-\frac{1}{\varepsilon}d\ln\sqrt{V}=-\frac{1}{2\varepsilon
V}V_{,I}\delta\varphi^I.
\end{equation}
Then it may be supposed that we can extract the derivation of $N$
with respect to $\varphi^I$,
\begin{equation}
\label{dNDPhi}N_{,I}\simeq-\frac{1}{2\varepsilon
V}V_{,I}=\frac{\varepsilon_I}{\sqrt{2}\varepsilon M_p}.
\end{equation}
However, this equation can not be applied to Eq.(\ref{FNL}) unless
the perturbations during multi-field inflation are purely
adiabatic. In fact, for a general multi-field model, the entropy
perturbation do exist. (See Ref.\cite{9511029} for an extensive
explanation.)

In order to use Eq.(\ref{dNDPhi}), in this section, we impose the
condition: \emph{Adiabatic Perturbations}. This implies that the
result in this section is only applicable to the multi-field
inflation during which the perturbations are purely adiabatic.

Then we get
\begin{equation}
\label{denominator}G^{IJ}N_{,I}N_{,J}=\frac{1}{2\varepsilon
M_p^2}.
\end{equation}
The derivation of $\varepsilon_I$ or $\varepsilon$ with respect to
$\varphi^J$ can be expressed approximately as
\begin{eqnarray}
\frac{\partial\varepsilon_I}{\partial\varphi^J}&\simeq&
-\frac{\partial}{\partial\varphi^J}\left(\frac{V_{,I}M_p}{\sqrt{2}V}\right)
=\frac{\sqrt{2}}{M_p}(\varepsilon_I\varepsilon_J-\frac{1}{2}\eta_{IJ}),
\label{dEpsilonLIDPhi}\\
\frac{\partial\varepsilon}{\partial\varphi^J}&=&
\frac{\partial}{\partial\varphi^J}(G^{IJ}\varepsilon_I\varepsilon_J)
\simeq{G^{KL}}_{,J}\varepsilon_K\varepsilon_L+
\frac{\sqrt{2}}{M_p}(2\varepsilon\varepsilon_J-G^{KL}\varepsilon_K\eta_{LJ}),
\label{dEpsilonDPhi}
\end{eqnarray}
with $\eta_{IJ}\equiv\frac{V_{,IJ}M_p^2}{V}$ and
${G^{KL}}_{,J}\equiv\frac{\partial G^{KL}}{\partial\varphi^J}$.
Now It is the time to calculate $N_{,IJ}$,
\begin{eqnarray}
N_{,IJ}&=&\frac{\partial}{\partial\varphi^J}\left(\frac{\varepsilon_I}{\sqrt2
\varepsilon}\right) \nonumber \\
&\simeq&\frac{1}{\varepsilon^2M_p^2}\{
G^{KL}\varepsilon_K\eta_{LJ}\varepsilon_I-
\frac{{G^{KL}}_{,J}}{\sqrt2M_p}\varepsilon_K\varepsilon_L\varepsilon_I-
\varepsilon\varepsilon_I\varepsilon_J-
\frac{1}{2}\varepsilon\eta_{IJ}\}.\label{dNDPhiDPhi}
\end{eqnarray}
Now it is easy to get
\begin{equation}
\label{numerator}G^{IK}G^{JL}N_{,I}N_{,J}N_{,KL}=
-\frac{1}{2\varepsilon^3M_p^3}{\beta},
\end{equation}
with
\begin{equation}
\label{beta}\beta\equiv\frac{\varepsilon^2}{M_p}+
\frac{{G^{MN}}_{,L}}{\sqrt2}G^{JL}\varepsilon_M\varepsilon_N\varepsilon_J-
\frac{1}{2M_p}G^{JL}G^{MN}\varepsilon_J\varepsilon_M\eta_{NL}.
\end{equation}

Substituting Eq.(\ref{denominator}) and (\ref{numerator}) into
Eq.(\ref{FNL}), we can express the nonlinear parameter in the form
as
\begin{equation}
\label{fNLSlowRollAppr}f_{NL}\simeq\frac{5}{3}\times\frac{\beta
M_p}{\varepsilon}
\end{equation}

Here we emphasize again that Eq.(\ref{fNLSlowRollAppr}) is only
applicable to the multi-field models in which the entropy
perturbations can be neglected.

\section{Summary}

In this paper, the ``$\delta N$-formalism" suggested to express
the non Gaussianity in \cite{a0504045} is generalized to the
multi-field inflationary models with the non-trivial target space
metric. One key step in our derivation is the equation
(\ref{spectrumOfField}). We believe that this equation is correct
\cite{a9507001,g9502002}. We have rewritten the result by using
the slow-rolling approximation, which is easy to be analyzed. But
this result is obtained under the condition of \emph{Adiabatic
Perturbations}. For a general multi-field model, we should use
Eq.(\ref{FNL}) to calculate $f_{NL}$.

Additionally, in this paper we have restricted our attention to
the contribution of the Gaussian part of $\delta\varphi^I$ and
ignored the contribution of the intrinsic non Gaussianity of
$\delta\varphi^I$, (\ref{intrinsicNG}). We emphasize that in some
case the term (\ref{intrinsicNG}) should be included. This will be
discussed in future work.

\end{document}